\documentclass[11pt, letterpaper]{article}
\usepackage[top=0.85in,left=1in,footskip=0.75in,marginparwidth=2in]{geometry}

\usepackage{amsmath}
\usepackage{amssymb}
\usepackage{amsthm}
\usepackage{cases}
\usepackage{siunitx}
\usepackage{verbatim}
\usepackage{newclude}
\usepackage{bm}
\usepackage[sort&compress,comma,super]{natbib}

\usepackage[utf8]{inputenc}

\usepackage{nameref,hyperref}

\usepackage{microtype}
\DisableLigatures[f]{encoding = *, family = * }

\usepackage{changepage}

\usepackage[aboveskip=1pt,labelfont=bf,labelsep=space,singlelinecheck=off]{caption}

\makeatletter
\renewcommand{\@biblabel}[1]{\quad#1.}
\makeatother

\usepackage{lastpage,fancyhdr,graphicx}
\usepackage{epstopdf}

\usepackage{color}

\definecolor{Gray}{gray}{.25}

\usepackage{graphicx}

\usepackage{sidecap}
\renewcommand{\figurename}{Fig.}
\usepackage{wrapfig}
\usepackage[pscoord]{eso-pic}
\usepackage[fulladjust]{marginnote}
\usepackage{siunitx}
\reversemarginpar

\begin{document}
\vspace*{0.35in}
\definecolor{bured}{rgb}{0.8, 0.0, 0.0}

\begin{flushleft}
{\Large
\textbf\newline{Learn to integrate parts for whole through correlated neural variability}
}
\newline
\\
Zhichao Zhu\textsuperscript{1,2}, 
Yang Qi\textsuperscript{1,2,3}, 
Wenlian Lu\textsuperscript{5,6,7,8},
Jianfeng Feng\textsuperscript{1,2,3,4,*}
\\
\bigskip
\it{1} Institute of Science and Technology for Brain-Inspired Intelligence, Fudan University, Shanghai, China
\\
\it{2} Key Laboratory of Computational Neuroscience and Brain-Inspired Intelligence (Fudan University), Ministry of Education, China
\\
\it{3} MOE Frontiers Center for Brain Science, Fudan University, Shanghai, China
\\
\it{4} Zhangjiang Fudan International Innovation Center, Shanghai, China
\\
\it{5} School of Mathematical Sciences, Fudan University, Shanghai, China
\\
\it{6} Shanghai Center for Mathematical Sciences, Shanghai, China
\\
\it{7} Shanghai Key Laboratory for Contemporary Applied Mathematics, Shanghai, China
\\
\it{8} Key Laboratory of Mathematics for Nonlinear Science, Shanghai, China
\bigskip

\end{flushleft}
\section*{Abstract}
Sensory perception originates from the responses of sensory neurons, which react to a collection of sensory signals linked to various physical attributes of a singular perceptual object. 
Unraveling how the brain extracts perceptual information from these neuronal responses is a pivotal challenge in both computational neuroscience and machine learning.
Here we introduce a statistical mechanical theory, where perceptual information is first encoded in the correlated variability of sensory neurons and then reformatted into the firing rates of downstream neurons.
Applying this theory, we illustrate the encoding of motion direction using neural covariance and demonstrate high-fidelity direction recovery by spiking neural networks.
Networks trained under this theory also show enhanced performance in classifying natural images, achieving higher accuracy and faster inference speed. 
Our results challenge the traditional view of neural covariance as a secondary factor in neural coding, highlighting its potential influence on brain function.
\section{Introduction}
A fundamental function of the brain is to encode and interpret the perceptual significance of stimuli encountered in the environment. 
Real-world stimuli, often complex and multifaceted, challenge sensory systems to integrate various characteristics into cohesive perceptual wholes, which retain crucial information for perception and behavior. 
Sensory neurons respond to specific stimulus attributes, such as light intensity, color, and sound frequency. However, individual neurons' responses, varying in spatial and temporal aspects, only provide partial insights into these perceptual wholes.
Consequently, spatial-temporal relationships among sensory neurons are pivotal for altering downstream neuronal activities, leading to successful perception~\cite{Ackels2021, Caruso2018, Borghuis2019, Panzeri2010}.
 
Understanding how downstream neurons process this sensory information is at the heart of computational neuroscience. 
The seminal work of Hubel and Wiesel~\cite{Hubel1962} introduced a feedforward model explaining the visual cortex's orientation selectivity in cats. 
Following this, a line of studies have utilized artificial nodes to model neural activities~\cite{Jones1987, Olshausen1996}. 
However, early research mainly used handcrafted parameters and simplified stimuli, thus unlikely to address complex perceptual tasks.
With great advances in deep learning~\cite{LeCun2015} and its historical ties with neuroscience~\cite{Hassabis2017}, there is an increasing trend to use artificial neural networks (ANNs) to understand neural computation in the brain~\cite{KhalighRazavi2014, Cichy2016, Li2023}.
However, a critical difference exists between ANNs and biological circuits in how they represent and propagate information. 
ANNs function by using continuous and deterministic activation values in noise-free environments. 
This allows for the availability of global information at every computational step.
In contrast, biological neurons communicate through discrete and irregular spiking activities~\cite{Tomko1974, Tolhurst1983, Softky1993}, which are temporally noisy but consistent in amplitude. 
This implies that biological neurons access only local information at a time. 
To capture temporal-distributed information, ANNs require structures such as recurrent connections with long short-term memory units or transformers, which are either not biologically plausible or non-causal~\cite{Li2023}. 
In comparison, biological neurons leverage their integrate-and-fire properties to process temporal information~\cite{Koenig1996}.
Thus, while conventional ANNs are valuable for understanding the 'what' of the brain's computations, they fall short in explaining the 'how'.

Addressing the computations undertaken by biological neurons requires understanding both how information is encoded in neuronal responses and how it affects downstream neurons~\cite{Panzeri2017}.
The firing rate, which is characterized as the number of spikes in a given time window, is a common method to encode information~\cite{Golledge2003, montani2007role, Rolls2011}.
However, increasing theoretical and experimental evidence indicates the importance of temporal structures and correlations between neuronal responses in representing information~\cite{Shamir2004, Averbeck2006a, Pillow2008, ElGaby2021, panzeri2022structures-reviews}.
By evaluating the disparity in information regarding the targeted perception offered by collective neuron responses as opposed to individual ones, we can assess the connection between the responses of sensory neurons and perceptual information~\cite{Pola2003, Schneidman2003}. 
When perception-related information is present in the collective responses but absent in individual responses, it indicates that such information arises from the synergistic coactivity of neurons, rather than from isolated neuronal responses.
Crucially, this synergistic effect is influenced by factors such as stimulus-induced synchrony or the rate at which spiking patterns occur~\cite{quiroga2013principles, panzeri2022structures-reviews}. 

Incorporating the perceptual disentangling hypothesis~\cite{DiCarlo2007, DiCarlo2012}, we introduce a novel concept of covariance-based computation in biological neurons. 
This computation is akin to a 'decorrelation' process, transitioning pertinent information from neuronal coactivities to the responses of individual neurons, thereby facilitating the linear decoding of perceptual information.
We start by introducing an encoding scheme that partitions sensory neuron covariance into components influenced by instantaneous stimulus intensity and neuronal activity fluctuations. 
We then demonstrate this scheme by encoding a moving grating's direction in the covariance of neurons that respond to light intensity and rate changes. 
Using the moment neural network (MNN) as the proxy, which rigorously characterizes spiking neural network (SNN) neuron statistics~\cite{Feng2006, Lu2010, Qi2023}, we can recover direction information under the principle of covariance-based computation.
After training, hidden neurons naturally exhibit direction selectivity akin to cortical neurons~\cite{Sclar1982, Kohn2005}, without the need for specific training constraints.
SNN simulations further show that hidden neurons decode motion direction from local fluctuations, bypassing the need for explicit global covariance information.
Information-theoretic analysis verifies that the desired information can be perfectly recovered. 
Moreover, networks employing our scheme enhance natural image classification accuracy and inference speed. 
Our findings challenge the perceived secondary role of neural covariance in neural coding, highlighting its significant influence on brain functions. 
\section{Results}

\subsection{Covariance-based neural coding and computational mechanisms}
\label{sec:task_overview}
\begin{figure}
    \centering
    \includegraphics[width=\linewidth]{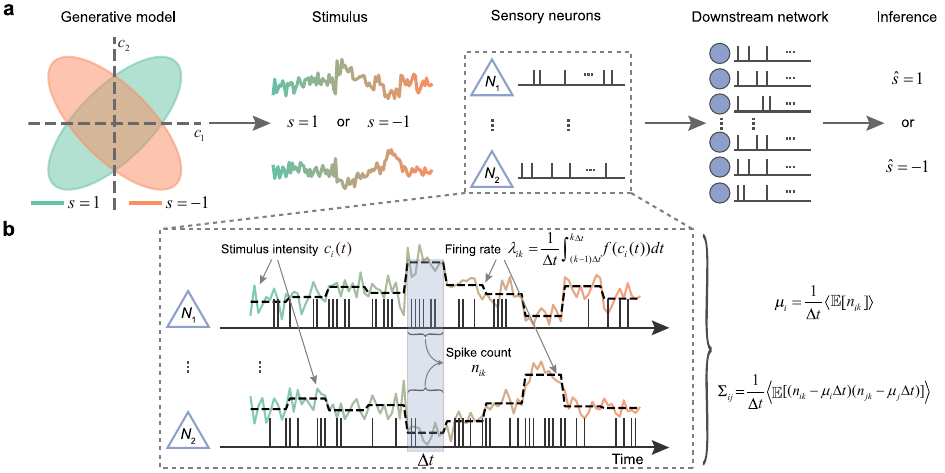}
    \caption{\textbf{Encode perceptual information in the spatial-temporal covariance of sensory neurons}.
   \textbf{a}, A schematic diagram showing how a stimulus \(s\) is represented by the correlated or anticorrelated appearance of two components \(c_1\) and \(c_2\).
   The intensity of each component reaching sensory neurons fluctuates, leading to variable neural activities.
   The downstream network estimates \(\hat s\) based on the responses of sensory neurons.
   \textbf{b}, A detailed examination of sensory neurons' activities reveals that the mean and covariance over short timescales capture the varying intensities of the stimulus components. The representation of stimulus \(s\) encoded in the sign of the correlation between sensory neurons \(N_1\) and \(N_2\), responsive to \(c_1\) and \(c_2\) respectively.}
   \label{fig:task_overview}
\end{figure}
To elucidate the concept of covariance-based neural computation, let us consider a binary decision-making task involving the identification of a stimulus \(s\) composed of two odor concentrations \(c_1\) and \(c_2\)~\cite{Ackels2021}. 
The co-release of these odors can be correlated or anticorrelated, contingent on the specific type of stimulus presented (\(s = 1\) or \(s = -1\)). 
In this scenario, an animal is trained to discriminate between these two types of stimuli.
While the mean concentrations \(c_1\) and \(c_2\) of the odors remain consistent irrespective of the stimulus \(s\), the actual concentrations reaching the animal's sensory neurons exhibit fluctuations due to turbulent air. 
Consequently, the animal must decipher the temporal correlation patterns of these fluctuating odor concentrations to accurately differentiate between the two stimuli.

We can conceptualize the stimulus reaching the sensory neurons as being drawn from a structured distribution, as depicted in Fig.~\ref{fig:task_overview}\textbf{a}.
This dynamic stimulus induces irregular spiking activities in sensory neurons. 
The downstream neurons are then tasked with interpreting the embedded information from these sensory neuron activity patterns, a crucial step in facilitating decision-making. 
This scenario prompts an essential inquiry: What specific aspects of sensory neuron activities are integral to conveying key information about the stimulus \(s\)?

We posit that information is naturally encoded through the covariance of temporally fluctuating stimuli. 
To illustrate the encoding of stimulus \(s\), we introduce two distinct types of sensory neurons, each tuned to component (\(c_1\) or \(c_2\)) of the stimulus, as shown in Fig.~\ref{fig:task_overview}\textbf{b}.
The response of these neurons is modeled as independent inhomogeneous Poisson processes with a rate function defined by
\begin{equation}
    \lambda_i(t) \doteq \lambda_{ik} = \frac{1}{\Delta} \int_{(k-1)\Delta t}^{k\Delta t}f(c_i(t))dt, \quad  (k-1)\Delta t \leq t < k \Delta t,
    \label{eq:rate_function_define}
\end{equation}
where \(c_i(t)\) represents the concentration of odor \(i\) at time \(t\), \(f\) is a function of the concentration and \(\Delta t\) the timescale over which sensory neurons can track changes~\cite{Panzeri2010, Theunissen1995}.

For a given time interval \(\Delta t\), both the mean and variance of the spike count \(n_{ik}\) in such a time interval \(\Delta t\) equal \(\lambda_{ik}\Delta t\).
Dividing the stimulus duration \(T\) into \(N_b = \lfloor T / \Delta t \rfloor\) bins allows for a piecewise constant representation of the mean firing rate within each bin \(k\).
Denoting the expected value of the spike count conditioned on the firing rate as \(\mathbb{E}[\cdot]\) and the average firing rate over index \(k\) as \(\langle \cdot \rangle\), we define the moments of random spike counts for a pair of neurons as
\begin{equation}
    \mu_i \equiv \frac{1}{\Delta t}\langle \mathbb{E}[n_{ik}] \rangle = \langle \lambda_{ik}\rangle,
    \label{eq:theory_define_mean}
\end{equation}
\begin{equation}
    \Sigma_{ij} \equiv \frac{1}{\Delta t}\langle \mathbb{E}[(n_{ik} - \mu_i\Delta t)(n_{jk} - \mu_j\Delta t)]\rangle  =  \Sigma^{\text{noise}}_{ij} + \Sigma^{\text{signal}}_{ij}.
    \label{eq:theory_define_cov}
\end{equation}
The first term on the right of equation~(\ref{eq:theory_define_cov}) is the noise covariance average over \(k\).
With the assumption that \(n_{ik}\) represent independent Poisson spikes, we get $\langle \Sigma_{ij\mid k}^{\text{noise}}\rangle = \delta_{ij}\lambda_i$,
where $\delta_{ij}$ represents Kronecker's delta, this simplifies into
\begin{equation}
    \Sigma_{ij}^{\text{noise}} = \delta_{ij}\mu_i.
    \label{eq:noise_cov}
\end{equation}
The second term corresponds to the cross-time signal covariance,
\begin{equation}
    \Sigma^{\text{signal}}_{ij} =  \langle(\lambda_{ik} - \mu_i)(\lambda_{jk} - \mu_j )\rangle \Delta t.
    \label{eq:signal_cov}
\end{equation}
The encoding scheme we describe exhibits several key characteristics.  
Both the mean firing rate and noise covariance, defined in equation~(\ref{eq:theory_define_mean}) and equation~(\ref{eq:noise_cov}) do not depend on the chosen timescale. 
In contrast, signal covariance is timescale-sensitive.
It diminishes when the stimulus is static~(\(\lambda_{ik} = \mu_i, \forall k \leq N_b\)) or when the observation window \(\Delta t\) is comparably large to the stimulus duration \(T\).
Similarly, if the stimulus fluctuates faster than the sensory neurons' tracking timescale, leading to firing rates that are uniformly in time [equation~(\ref{eq:rate_function_define})], signal covariance will also be negligible.
Unlike using raw firing sequences, our statistical encoding scheme is not affected by changes in stimulus intensity sequence, but by correlations in neuronal firing rates. This approach ensures a consistent perceptual representation, regardless of stimulus intensity variability.

The perceptual information about \(s\) cannot be discerned by observing \(c_1\) or \(c_2\) in isolation, but it becomes evident when considering the correlation between these two components, which is reflected in the correlated neural variability observed in sensory neurons. 
In the perspective of neural computation, the irregular spike trains of presynaptic neurons induce fluctuating synaptic currents, leading to irregular spiking in postsynaptic neurons~\cite{de2007correlation}. 
Under the framework of MNN, this stochastic computation can be elegantly captured by moment mapping~\cite{Qi2023}.
The covariance-based computation in this context is characterized by learning a nonlinear mapping \(\phi\), which converts the mean and covariance of the inputs into specific outputs, \(\mu', \Sigma' = \phi(\mu, \Sigma)\). 
This transformation aims to effectively shift the perceptual information that is initially encoded in the covariance \(\Sigma\) to the firing rates \(\mu'\) of downstream neurons, facilitating its interpretation and guiding decision-making. 
\subsection{Encoding motion directions through neural covariance}
\label{sec:vo_encode}
In this section, we explore the covariance-based encoding scheme using a motion direction detection task, a common method for studying early visual information processing. 
The task involves showing a subject a moving visual grating, oriented perpendicularly to its motion direction (illustrated in Fig.~\ref{fig:visual_orientation_mnn}\textbf{a}, leftmost panel), and having them identify its movement direction.
Our goal is to demonstrate the effective encoding of motion direction within the temporal correlations of sensory neuron responses, which respond to basic physical properties such as light intensity and its changing rate.

The intensity of a moving grating stimulus is described by a sinusoidal function,
\begin{equation}
    I(\mathbf{x}, t) = 1 + c \cos(\mathbf{k \cdot x} - \omega t)
\end{equation}
where $c \in [0, 1]$ is the contrast, \(\omega\) is the temporal angular frequency, 
and \(\mathbf{k}\) is the spatial wave vector, indicating the motion direction. 
The length of \(\mathbf{k}\), denoted as \(k = |\mathbf{k}|\), represents the grating's spatial frequency. 
We model the response of sensory neurons as an inhomogeneous Poisson process with a rate function \(\lambda_i(t) = \alpha I(x_i, t)\), where \(x_i \in \mathbb{R}^2\) is the spatial location of the neuron and \(\alpha\) acts as a gain factor, modulating the neuron's response to the stimulus intensity.
Assuming the rate function variation to be slow relative to the observation window \(\Delta t\), the spike count $n_i$ within this interval has statistical moments given by:
\begin{equation}
    \mathbb{E}[n_i] = \alpha \Delta t,
    \label{eq:mean_n}
\end{equation}
\begin{equation}
    {\rm Cov}[n_i, n_j] =  \alpha \delta_{ij}\Delta t + \tfrac{1}{2}\alpha^2 c^2
    \cos[\mathbf{k} \cdot (\mathbf{x}_i - \mathbf{x}_j)]\Delta t^2.
    \label{eq:cov_n}
\end{equation}
Here, the mean intensity encodes global luminance, while contrast and spatial direction are captured in the covariance, with the spatial direction being fully encoded by the correlation coefficients. 
However, intensity encoding alone cannot distinguish between opposite motion directions \(\mathbf{k}\) and \(-\mathbf{k}\), as they yield identical covariance values.
To address this, we introduce an additional input channel based on the change rate of intensity:
\begin{equation}
    \partial_t I(\mathbf{x}, t) = 1 - c\omega \sin(\mathbf{k \cdot x} - \omega t).
\end{equation}
This change detection is also modeled as an inhomogeneous Poisson process with a rate function $\beta \partial_t I(x_k, t)$,
where $x_k$ are the spatial locations of the change detectors and 
$\beta$  serves as the gain factor for the neural response to intensity changing rate.
The statistical moments for these change detectors over $\Delta t$ are
\begin{equation}
    \mathbb{E}[n_k] = \beta \Delta t,
\end{equation}
\begin{equation}
    {\rm Cov}[n_k, n_l] = \tfrac{1}{2} \beta^2 c^2 \omega^2
\cos[\mathbf{k} \cdot (\mathbf{x}_k - \mathbf{x}_l)] \Delta t^2 + \beta\delta_{kl} \Delta t .
\end{equation}
Notably, the variance now incorporates information about the temporal frequency $\omega$.
New motion direction information emerges in the correlation between intensity and its rate of change:
\begin{equation}
    {\rm Cov}[n_i, n_k] 
    = -{\rm Cov}[n_k, n_i]
    = \tfrac{1}{2} \alpha \beta c^2 \omega\sin[\mathbf{k} \cdot (\mathbf{x}_i - \mathbf{x}_k)]\Delta t^2.
    \label{eq:motion_information_embed}
\end{equation}
Importantly, this representation of the moving direction in the covariance is independent of the initial spatial phase of the intensity and change detectors, underscoring the robustness and phase-invariance of the encoding scheme.

To specify the input layer of our neural network, we consider a hexagonal grid with \(N\) sites, where each site hosts one intensity detector \(i\) and one change detector \(k\).
The inputs to the MNN can therefore be compacted in matrix form as 
\begin{equation}
    \mu = \tfrac{1}{\Delta t}(\mathbb{E}[n_i], \mathbb{E}[n_k])^T
    \label{eq:vo_input_mean}
\end{equation}
and
\begin{equation}
\Sigma = \tfrac{1}{\Delta t} \begin{pmatrix}
{\rm Cov}[n_i, n_j], {\rm Cov}[n_i, n_l] \\
{\rm Cov}[n_k, n_j], {\rm Cov}[n_k, n_l]
\end{pmatrix} \label{eq:vo_input_cov}
\end{equation}
Note that our method also works if the input neurons are scattered randomly in space.

Before applying this input to MNN, we must consider a caveat.
MNN defines covariance for stationary processes in the infinite time window limit, while the covariance in equation (\ref{eq:theory_define_cov}) is computed for point processes riding on slowly varying rates over a finite time window.
This violates both the stationarity and the limit conditions, potentially causing deviations between the two definitions of covariance.
To reconcile this, we propose preserving stationarity by ensuring that the rate varies slowly within one observation time window and collecting a sufficiently large number of spikes over the finite time window to approximate the theoretical limit.
While this encoding scheme is a simplified model, it serves as a valuable conceptual framework, highlighting the significance of covariance in neural computation. 
In the forthcoming section, we will demonstrate how the downstream network is trained to decode motion directions 
\subsection{Learning to decode motion directions through covariance-based computation}
\label{sec:vo_mnn}
\begin{figure}
    \centering
    \includegraphics[width=\linewidth]{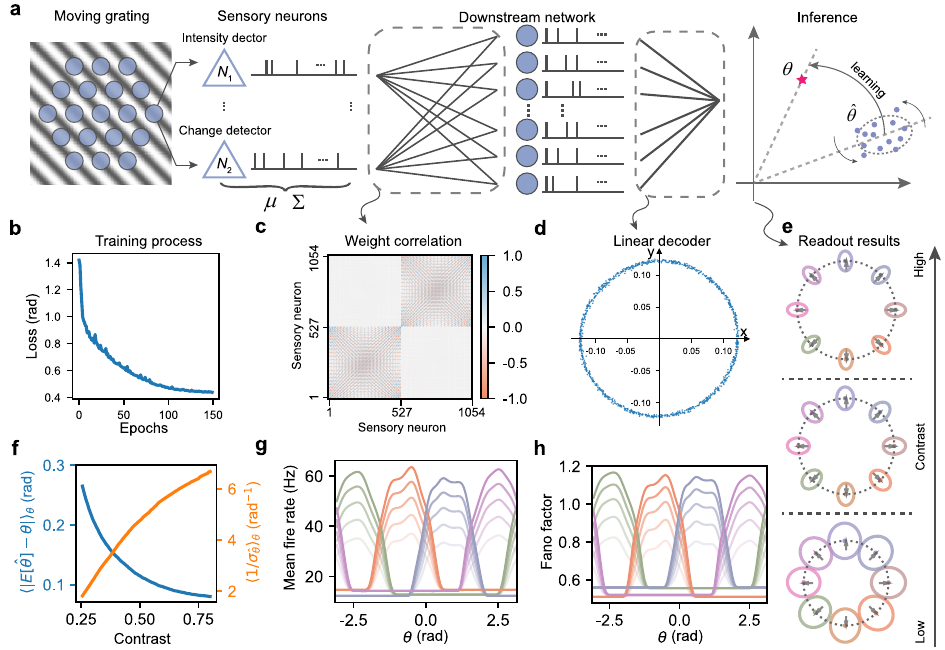}
    \caption[visual orientation mnn]{\textbf{Learn to infer motion direction through correlated neural variability.}
    \textbf{a}, Task overview: Neurons arranged in a hexagonal grid collectively respond to the intensity of light and the corresponding rate of change of a moving grating.
    The downstream network learns to have a better estimate by minimizing readout errors and trial-to-trial variability.
    \textbf{b}, Evolution of validation loss across training epochs.
    \textbf{c}, Correlation pattern of the trained weights \(W_{\rm in}\) for each sensory neuron projecting to hidden neurons.
    Indexes 1-527 denote intensity detectors, while the remaining represent change detectors.
    \textbf{d}, Visualization of the trained linear decoder \(W_{\rm out}\), where the X and Y coordinates of each dot are the connection strengths that map the response of each hidden neuron to the 2D readout space.
    \textbf{e},  Normalized readout results for various motion directions at different contrast levels (\(c \in [0.25, 0.5, 0.75]\)). 
    Gray arrows represent ground truths, while colored dots and ellipses illustrate readout mean and covariance, respectively. 
    \textbf{f}, Average readout error (blue) and readout variability (orange) as a function of contrast.
    \textbf{g-h},  Tuning curves and Fano factor curves of hidden neurons with specific preferred motion directions (indicated by color, as in \textbf{e}). Transparency corresponds to contrast levels (higher contrast results in lower transparency).}
    \label{fig:visual_orientation_mnn}
\end{figure}
In our motion direction task, we employ a feedforward MNN depicted in Fig.~\ref{fig:visual_orientation_mnn}\textbf{a}, consisting of sensory neurons, a single hidden layer, and a linear readout. 
The inputs include the mean and covariance of the sensory neuron responses as defined by Eq~(\ref{eq:vo_input_mean})-(\ref{eq:vo_input_cov}).
Using moment mapping~\cite{Qi2023}, the first and second moments of the responses of the hidden neurons are 
\begin{equation}
    \label{eq:input2hidden}
    (\mu', \Sigma') = \phi(W_{\rm in}\mu + \mu_{\rm ext}, W_{\rm in}\Sigma W_{\rm in}^T + \Sigma_{\rm ext}),
\end{equation}
where \(\phi\) is the moment activation function~\cite{Qi2023} [equation~(\ref{eq:ma_mu})-(\ref{eq:ma_chi}) in Methods], and \((\mu_{\rm ext}, \Sigma_{\rm ext})\) depict the moments of external currents. 
Here, the mean external current $\mu_{\rm ext}$ is a trainable parameter, while the covariance \(\Sigma_{\rm ext}\) is fixed at 0 mA/ms.

As the mean firing rates of sensory neurons do not reflect the motion directions and linear transformations cannot alter the linear Fisher information, the nonlinear coupling through moment activation \(\phi\) is crucial for extracting information from covariance,
The hidden neurons' responses are then mapped to a 2D vector for the estimated direction (Fig.~\ref{fig:visual_orientation_mnn}\textbf{a}). 
The mean and covariance of the readout are given by
\begin{equation}
    (\hat{\mu}, \hat{\Sigma}) = (W_{\rm out}\mu', W_{\rm out}\Sigma' W_{\rm out}^T),
\end{equation}
corresponding to the distribution of possible estimated direction angle \(\hat \theta\) (the dots in the rightmost panel), while the ground truth \(\theta\) is marked as a red star with coordinates (\(\cos \theta, \sin \theta\)). 

To train the network, we introduce a loss function targeting both the average and variability discrepancies between the estimated and true directions [equation~(\ref{eq:loss_function})].
The network is trained on a dataset of gratings moving in diverse directions for 150 epochs, showing a decreasing validation loss that converges after about 100 epochs (Fig.~\ref{fig:visual_orientation_mnn}\textbf{b}).

After training, we first assess the model parameters. 
We analyze the influence of sensory neurons on hidden neurons by calculating the column-wise correlation of synaptic weight \(W_{\rm in}\). 
A higher correlation suggests more similar effects of sensory neurons on hidden neurons. 
As shown in Figure~\ref{fig:visual_orientation_mnn}\textbf{c}, synaptic weights are typically correlated or anticorrelated based on spatial positions and type (intensity or change detectors).
Correlations between intensity and change detectors are minimal, indicating their independent roles in information transmission. 
The linear decoder \(W_{\rm out}\), illustrated in Figure~\ref{fig:visual_orientation_mnn}\textbf{d}, displays a ring-shaped structure, with each dot's coordinates denoting the readout weight from a hidden neuron. 
This structure reflects the direction selectivity of hidden neurons, encoding direction information in their mean firing rates for linear decoding.
 
We then evaluate the model performance by varying stimulus contrasts and directions.
Higher contrast levels show greater alignment of the mean readout (dots in Fig.~\ref{fig:visual_orientation_mnn}\textbf{e}) with the ground truth (gray arrow), reducing the average discrepancy (Fig.~\ref{fig:visual_orientation_mnn}\textbf{f}, blue line). 
The readout covariance for all stimuli forms a radial pattern (ellipse in Fig.~\ref{fig:visual_orientation_mnn}\textbf{e}), with principal axes parallel to the readout mean, minimizing random errors from input noise.
Lower contrasts lead to expanded, less eccentric covariance, increasing estimate variability. 
This results in the variance of the estimated angle \(\hat\theta\) being inversely related to stimulus contrast (Fig.~\ref{fig:visual_orientation_mnn}\textbf{f}, orange line), consistent with the predictions under probabilistic population code theory~\cite{Ma2006}.

We further explore the activity characteristics of hidden neurons by analyzing their tuning functions and Fano factors.
Figure \ref{fig:visual_orientation_mnn}\textbf{g} shows four hidden neurons with bell-shaped tuning curves.
Each has a preferred direction and the height varies with contrast, but the width remains consistent.
Noise correlation among hidden neurons decreases with greater differences in preferred directions and correlates with the similarity in connections to sensory neurons (Extended Data Fig.~\ref{fig_s1: correlation_vs_preferred_direction}).
These features are akin to those in direction-selective cortical neurons~\cite{Sclar1982, Kohn2005}, achieved without specific training constraints.
The hidden neurons' Fano factor curves, resembling their mean firing rate profiles, peak at preferred directions and amplify with increased contrast(Fig.~\ref{fig:visual_orientation_mnn}\textbf{h}). 
This reflects the encoding of motion direction in covariance, where higher contrast boosts the signal covariance, leading to a higher response variability of hidden neurons.
Contrary to the notion that noisier neuron activity hampers coding efficiency, our results show improved readout accuracy at higher contrasts.

Overall, our findings stem from three sources of hidden neuron responses outlined in equation~(\ref{eq:input2hidden}): a constant external current \(\mu_{\rm ext}\), activity fluctuations with direction-independent statistical properties \(\mu\) and \(\Sigma_{\rm noise}\), and direction-dependent variation \(\Sigma_{\rm signal}\), where \(\Sigma = \Sigma_{\rm noise} + \Sigma_{\rm signal}\). 
The first two elements are not specific to direction, whereas the third, influenced by trained weights, is the key to direction selectivity. 
With increased contrast, the correlation between intensity and change detectors strengthens, eliciting more pronounced stimuli-related responses in hidden neurons and elevating their Fano factors.
\subsection{Dynamics and efficacy of covariance-based computation in SNNs}
\label{sec:vo_snn}
\begin{figure}
    \centering
    \includegraphics[width=\linewidth]{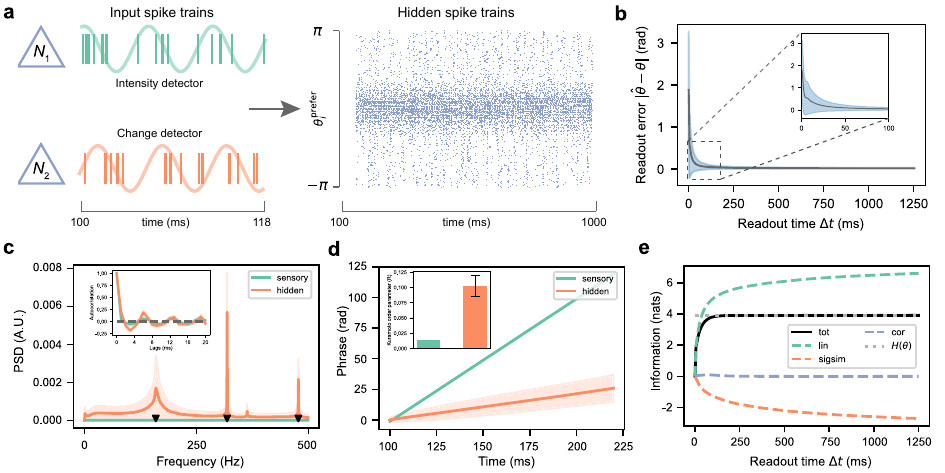}
    \caption[visual orientation snn]{\textbf{Dynamics of SNN in integrating local information for decoding directions}.
    \textbf{a}, Raster plots that represent an intensity detector and a change detector at the same spatial location, alongside hidden neurons of the SNN under a moving grating with a direction of \(\theta = 0.06\) rad. 
    The hidden neurons are arranged based on their preferred directions.
    \textbf{b}, Readout error of the SNN as a function of readout time \(\Delta t\). 
    The inset provides a detailed view of the initial 100 ms of the readout error curve.
    \textbf{c}, Power spectral density and autocorrelation (inset) of the relative deviation of population spike count (1 ms time window).
    Black triangles mark the temporal frequency of the stimulus, including double and triple frequencies.
    \textbf{d}, Unwrapped average phase and average Kuramoto order parameter (inset) of sensory neurons' firing rates and hidden neurons' membrane potentials from 100 to 220 ms after stimulus onset.
    \textbf{e}, Quantification of direction information in the readout. The gray dotted line represents the theoretical bound of mutual information, specifically the entropy of moving direction  \(H(\theta)\). 
    Components: \textit{tot} - total mutual information between stimuli and readouts; \textit{lin} - the sum of mutual information from individual neuron responses; \textit{sigsim} - the difference in entropy between population and individual responses; \textit{cor} - information from correlated neuron activities.
    In \textbf{b}, \textbf{c}, \textbf{d}, solid lines and shaded regions (with an error bar in the inset of \textbf{d}) depict the mean and standard deviation over 500 trials, averaged across 50 motion directions.}
    \label{fig:visual_orientation_snn}
\end{figure}
So far we have modeled covariance computation by using the MNN model to explicitly track the neural pair-wise covariance.
At first glance, this process may necessitate global knowledge of the covariance structure of the input. 
However, this can be done implicitly by the stochastic process of neural spikes, and each sensory neuron only has access to local information. 
To illustrate this, we simulate a spiking neural network to show how the information hidden in the covariance structures of the input can be extracted without explicitly knowing the covariance.

Derived analytically from the leaky integrate-and-fire model,  we can use the trained MNN to reconstruct the SNN without additional tuning. 
Figure~\ref{fig:visual_orientation_snn}\textbf{a} displays the spike trains from a pair of sensory neurons and neurons from the hidden layer. 
The two sensory neurons, located in the same spatial location, detect light intensity and its changing rate using an inhomogeneous Poisson process with oscillatory firing rates (green and orange curves, left panel), as detailed in Section~\ref{sec:vo_encode}.
Their firing rates range from 200 to 1800 spikes per second, under the stimuli contrast \(c=0.8\). 
In contrast, hidden neurons exhibit sparser firing patterns, ranging from 0 to 200 spikes per second as shown in the raster plot (right panel), with those whose preferred directions are closer to the stimulus direction displaying notably higher firing rates.
Using the MNN's trained linear decoder on hidden neurons' spike trains leads to increasingly precise direction estimates. 
Figure~\ref{fig:visual_orientation_snn}\textbf{b} demonstrates a decrease in both the mean and variability of readout error with longer \(\Delta t\). 
The mean error converges near zero at about 50 ms, while the standard deviation stabilizes close to zero after 100 ms.

To compare the differences in spike activities between sensory and hidden neurons, we analyze the power spectral density of their relative deviation of spike count at the population level, using 1 ms time windows. 
Figure~\ref{fig:visual_orientation_snn}\textbf{c} reveals that the power spectral density for sensory neurons is nearly zero, suggesting that the distinct instantaneous firing rates of individual sensory neurons counterbalance each other at the population level, resulting in stable overall activity.
Conversely, the power spectral density of hidden neurons peaks significantly at multiples of 159 Hz, reflecting the stimulus's temporal frequency with notably higher amplitudes than those of sensory neurons.
The greater amplitudes in the population spike count's relative deviation suggest that hidden neurons display more oscillatory activity than sensory neurons.
This results from the \(\pi/2\) phase difference between the firing rates of intensity and change detectors, creating a time delay in spike waves between these neuron groups. 
Coupled with correlated weights (Fig.~\ref{fig:visual_orientation_mnn}\textbf{c}), this enhances the likelihood of collective neuronal activity triggering hidden neurons' firing.
Furthermore, the autocorrelation of relative deviation (Fig.~\ref{fig:visual_orientation_snn}\textbf{c} inset) shows sensory neuron activity oscillating at about 1.0 rad/ms, aligning with the stimulus frequency but weakly correlated. 
Hidden neurons, in contrast, display stronger correlations than sensory neurons at short time lags, with both neuron types' correlation coefficients converging at longer lags.

Similar findings are obtained by analyzing sensory and hidden neurons' activities using the mean phase and Kuramoto order parameter (Fig.~\ref{fig:visual_orientation_snn}\textbf{d}). 
These metrics effectively assess phase shift orientation and synchronization levels. 
Our analysis, centering on the firing rate of sensory neurons and the membrane potential of hidden neurons, benefits from these continuous variables for accurate measurement and evaluation.
We find that the average phases of both sensory and hidden neurons linearly increase over time, mirroring the stimulus's linear phase progression. 
This implies that both the stimulus phrase and the sensory neurons' phrase can be inferred from the hidden neurons' phase.
Regarding the Kuramoto order parameter, sensory neurons exhibit an average below 0.025 (see Fig.~\ref{fig:visual_orientation_snn}\textbf{d} inset), indicating asynchronous firing patterns. 
In contrast, hidden neurons demonstrate a higher average parameter, approximately 0.1, signifying more pronounced oscillatory activity than sensory neurons.

To evaluate our information transition process, we quantify the amount of recovered direction information and its contributors through an information-theoretical analysis~\cite{Pola2003}. 
This analysis decomposes the mutual information \(I_{\rm tot}\) between readouts and motion directions into three components: \(I_{\rm lin}\), the information from each readout dimension separately; $I_{\rm sigsim}$, the redundancy due to the correlated responses shared in all readout dimensions; and $I_{\rm cor}$, the residual information from the correlations within the readout.
Figure~\ref{fig:visual_orientation_snn}\textbf{e} shows a rapid initial increase in \(I_{\rm tot}\) within the first 100 ms, eventually matching the entropy of moving directions \(H(\theta)\) and indicating lossless transmission of direction information. 
The primary contributor to \(I_{\rm tot}\) is \(I_{\rm lin}\), signifying that direction information is predominantly encoded in the firing rates of hidden neurons. 
While \(I_{\rm lin}\) continues to rise beyond 100 ms, its growth is counterbalanced by \(I_{\rm sigsim}\), keeping \(I_{\rm tot}\) steady. 
This redundancy is natural, as knowledge of one coordinate in a fixed direction vector restricts the possibilities for the other coordinate. 
With increasing readout time windows, the readout variance decreases but the mean remains constant, leading to more redundancy. 
In contrast, the correlation component \(I_{\rm cor}\) contributes minimally to direction information, remaining close to zero.

Our results are in agreement with both theoretical and experimental research~\cite{Schneidman2003, Averbeck2006, Golledge2003, montani2007role, Kriegeskorte2021}, which indicate that neural correlations have limited impact on information coding when stimuli are distinctly discerned by uncorrelated neuronal responses. 
Given that the maximum mutual information is limited by the stimuli's entropy, and considering the data processing inequality which states that downstream information cannot exceed upstream information, the importance of neural correlation in information transmission diminishes as the information encoded in a neuron's response becomes more pronounced. 
Therefore, the process of representational disentanglement can be interpreted as a 'decorrelation' process, wherein essential information shifts from the collective structure of the neuron population to individual neurons.
\subsection{Enhancing performance on natural image classification with covariance-based computation}
\label{sec:fine_grain_task}
\begin{figure}
    \centering
    \includegraphics[width=\linewidth]{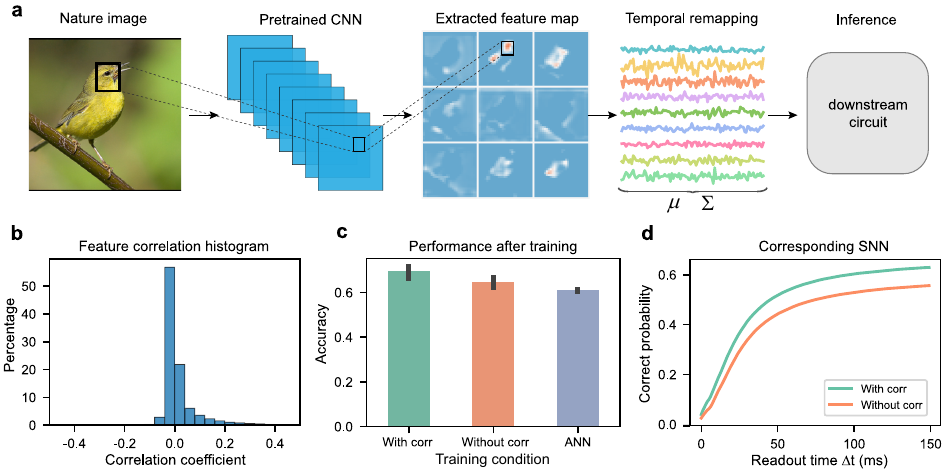}
    \caption{\textbf{Incorporating second-order information improves model performance in natural image classification}
    \textbf{a}, Task schematic: A pre-trained convolutional neural network (CNN) serves as the sensory system, extracting diverse features from the input image across multiple channels. These spatial features are then transformed into responses in the temporal domain, and their mean and covariance are computed using equation~(\ref{eq:theory_define_mean})-(\ref{eq:theory_define_cov}). 
    The downstream network utilizes the mean and covariance information to infer the image's category.
    \textbf{b}, Distribution of correlation coefficients in the covariance matrix obtained from all images in the dataset.
    \textbf{c}, Comparison of the classification accuracy of MNNs (\textit{With corr} and \textit{Without corr}) and an ANN after training. The error bars represent the standard deviation of 5 trials.
    \textbf{d}, The average probability of the correct prediction of the SNNs as a function of readout time.
    \textit{With corr}, an MNN trained with mean and covariance.
    \textit{Without corr}, an MNN trained with mean and shuffled covariance (all off-diagonal correlation coefficients set to zero).
    \textit{ANN}, an ANN that uses ReLU activation for comparison, trained with mean only.}
    \label{fig:fine_grain}
\end{figure}
We expand our covariance-based computation framework to tackle a more challenging task involving complex visual stimuli, moving beyond elemental features like motion direction. 
For this purpose, we focus on a fine-grained classification task involving natural bird images, which includes multiple subordinate categories within the broader category of birds~\cite{WahCUB_200_2011}. 
The complexity of the task arises from the subtle distinctions between classes and significant intraclass variations~\cite{Wei2022}. 
Furthermore, this task exemplifies a perceptual disentangling challenge, as natural images possess intricate structures where the interrelations between image patches (stimuli parts) are crucial in defining the object category (perceptual whole).

We explore the potential enhancement of classifier performance through the inclusion of covariance information from feature maps during training. 
These feature maps are generated from a natural image using a pretrained convolutional neural network (CNN)~\cite{simonyan2014very}, serving as the sensory system for image encoding. 
The CNN's output consists of various neural responses, each channel representing a detector for specific features, such as a bird's beak. 
To use these data, we sequentially remap the spatial feature map to temporal data, representing instantaneous firing rates \(\lambda(t)\)of an inhomogeneous Poisson process. 
This approach allows us to calculate the mean and covariance of these feature maps, following the methodology described in equation~(\ref{eq:theory_define_mean})-(\ref{eq:theory_define_cov}). 

Before training the classification model, we analyze the distribution of off-diagonal entries in the correlation matrices of CNN-derived feature maps (Fig.~\ref{fig:fine_grain}\textbf{c}). 
Most correlation coefficients are found to lie within the \(\pm 0.2\) range, averaging at 0.0046. 
To assess whether these weak pairwise correlations hold significant information pertinent to image categorization, we feed the statistical moments of the CNN's feature map to a two-layer classifier, examining the impact of correlation on performance through three distinct models. 
The first model, termed \textit{With corr}, is an MNN trained with the mean and covariance of CNN's feature map directly.
The second model, \textit{Without corr}, is also an MNN but omits input correlations by zeroing the off-diagonal coefficients, effectively negating their influence while retaining the variance. 
The third model, \textit{ANN}, employs an ANN with rectified linear unit (ReLU) activation, trained exclusively on the mean firing rate.
To ensure fair and accurate comparisons, we maintained consistent network architectures and loss functions in these different setups.

Each model is trained 5 times with different weight initialization.
The \textit{ANN} model, which is based solely on the mean of the feature maps, demonstrates the lowest accuracy at 61\% on the test set. 
This suggests that relying exclusively on the mean is insufficient for optimal classification. 
The \textit{Without corr} model, incorporating individual neuron variances in training, shows a marginal improvement with an accuracy of 64.77\%. 
Further evaluation is conducted on SNN models reconstructed from the MNN models. 
These SNNs are given inputs that are taken from a multivariate normal distribution that has the same mean and covariance as the inputs of the MNNs.
Fig.~\ref{fig:fine_grain}\textbf{d} illustrates that the \textit{With corr} model, which accounts for correlation, consistently shows faster improvement and a higher likelihood of accurate predictions compared to the \textit{Without corr} model, which disregards input correlations.

In conclusion, extending our covariance-based computation theory to the realm of natural image classification has highlighted the significant benefits of incorporating correlations for enhanced task performance. 
Despite the typically weak nature of these correlation coefficients, they contain nuanced, category-specific information that augments the information provided by the mean.
This inclusion of correlations not only leads to higher classification accuracy but also improves the SNN's inference speed. 
These findings resonate with similar observations in the field of deep learning~\cite{Wang2020, Song2022}, further validating the importance of considering correlations in neural network training.
\section{Discussion}
In this study, we proposed a novel perspective on how perceptual information is embedded in correlated neural variability and how this covariance gets involved in neural computation.
We introduced an encoding scheme using the covariance of sensory neurons to capture complex perceptual wholes, a significant advance considering neurons' stochastic activities in the time domain.
Through a motion direction detection task, we showed that downstream networks are capable of learning covariance-based computation to decode this encoded information effectively. 
This process allows for the transformation of perceptual information from the covariance observed in upstream neurons to the firing rates of downstream neurons, based on the perceptual disentangling hypothesis~\cite{DiCarlo2007, DiCarlo2012}. 
Additionally, we demonstrated that SNNs could perform this computation implicitly, leading to lossless recovery of direction information. 
Moreover, our approach proved advantageous in a complex natural image classification task, enhancing model performance and inference speed, underscoring the significance of correlated neural variability in sophisticated information processing.

Our encoding scheme hinges on two essential components: a variety of neurons responsive to different stimulus aspects and an appropriate observation window to track stimulus-induced changes in neuronal responses. 
This approach aligns with the concept of a combinatorial code~\cite{Malnic1999, Osborne2008}, where neural responses at different timescales represent different stimulus characteristics~\cite{Panzeri2010}.
The resulting activity patterns of neural populations, shaped by stimuli, form a consistent neural manifold~\cite{DiCarlo2007, DiCarlo2012}, which remains stable despite the variability of the trial-to-trial response, as encapsulated by the covariance in our scheme.
From the decoding perspective, downstream neurons serve dual roles as integrators and coincidence detectors~\cite{Koenig1996}, because processing perceptual information requires extended timescales and collaboration between sensory neurons. 
We describe this process as covariance-based computation, where the covariance of sensory neuron spike trains dominates the firing rates of downstream neurons.
This finding suggests a hierarchical structure in neural processing, achieved through iterative covariance-based computations. 
As one ascends this hierarchy, the timescale and complexity of the encoded information increase, leading to a progressively linear representation of perceptual information~\cite{piasini2021temporal, Siegle2021}.

While our results are promising, we must acknowledge certain limitations. 
The downstream networks were trained using gradient-based backpropagation, capable of explicitly computing gradients related to covariance. 
However, the exact learning mechanisms of the brain, especially akin to backpropagation, are still not fully understood.  
Further research could delve deeper into the effects of local learning rules that are impacted by correlated neural activities.
For example, employing a burst-dependent synaptic plasticity rule~\cite{Payeur2021} could modify error signals in feedback connections. 
Such exploration could shed light on the prevalence of noise correlation in the brain and its potential role in enhancing learning processes.

The approach developed in this work also has broader implications for SNN training. 
We have demonstrated the benefits of incorporating second-order information, such as covariance, into model training. 
Integrating covariance into ANNs requires a quadratic number of parameters, making it difficult to use for larger models~\cite{Wang2020}. 
In comparison, SNNs can unfold the covariance over time, which allows them to implicitly process the correlation between variables, thus improving performance without the need for extra parameters.
Ultimately, our work paves the way for novel studies on the roles of noise and neural correlation in the brain, as well as the development of powerful learning algorithms for SNNs.
\section{Methods}
\subsection{Leaky integrate-and-fire neuron model}
We employed the leaky integrate-and-fire (LIF) spiking neuron model to describe the membrane potential dynamics of neurons:
\begin{equation}
    \dfrac{dV_i}{dt}= -LV_i(t) + I_i(t),
    \label{eq:LIF}
    \end{equation} 
where the sub-threshold membrane potential $V_i(t)$ of a neuron $i$ is driven by the total synaptic current $I_i(t)$ and $L=0.05$ \si{\per\milli\second} is the leak conductance. 
When the membrane potential $V_i(t)$ exceeds a threshold $V_{\rm th}=20$ \si{\milli\volt} a spike is emitted, as represented by a Dirac delta function. Afterward, the membrane potential $V_i(t)$ is reset to the resting potential $V_{\rm res}=0$ mV, followed by a refractory period $T_{\rm ref}=5$ ms. The synaptic current takes the form
\begin{equation}
    I_i(t)= \sum_{j}w_{ij}S_j(t)+I_i^{\rm ext}(t),
    \label{eq:current}
\end{equation}
where $S_j(t)=\sum_k \delta(t-t^k_j)$ represents the spike train generated by presynaptic neurons.

A final output $\mathbf{y}$ is readout from the spike count $\mathbf{n}(\Delta t)$ of a population of spiking neurons over a time window of duration $\Delta t$ as follows
\begin{equation}
y_i(\Delta t) = 
\frac{1}{\Delta t}\sum_{j} w_{ij}n_{j}(\Delta t) + \beta_i, 
\label{eq:snn_readout}
\end{equation}
where $w_{ij}$ and $\beta_i$ are the weights and biases of the readout, respectively. 
A key characteristic of the readout is that its variance decreases as the time window $\Delta t$ increases.

\subsection{Moment embedding for the leaky integrate-and-fire neuron model}
The moment embedding approach~\cite{Feng2006, Lu2010, Qi2023} begins with mapping the fluctuating activity of spiking neurons to their respective first- and second-order moments
\begin{equation}
\mu_i = \lim_{\Delta t\to\infty} \dfrac{\mathbb{E}[n_i(\Delta t)]}{\Delta t}, 
\label{eq:def_mean}
\end{equation}
and
\begin{equation}
\Sigma_{ij} = \lim_{\Delta t\to\infty} \dfrac{{\rm Cov}[n_i(\Delta t),n_j(\Delta t)]}{\Delta t}, 
\label{eq:def_cov}
\end{equation}
where $n_i(\Delta t)$ is the spike count of neuron $i$ over a time window $\Delta t$. 
In practice, the limit of $\Delta t\to\infty$ is interpreted as a sufficiently large time window relative to the timescale of neural fluctuations. 
We refer to the moments $\mu_i$ and $\Sigma_{ij}$ as the mean firing rate and the firing covariability in the context of MNN, respectively. 

For the LIF neuron model [equation (\ref{eq:LIF})], the statistical moments of the synaptic current are equal to~\cite{Feng2006, Lu2010}
\begin{numcases}{}
\hat{\mu}_i = \textstyle\sum_kw_{ik}\mu_k + \hat{\mu}_i^{\rm ext},\label{eq:sum_mean}
\\
\hat{\Sigma}_{ij}=\textstyle\sum_{kl} w_{ik}C_{kl}w_{jl} + \hat{C}_{ij}^{\rm ext},\label{eq:sum_cov}
\end{numcases}
where $w_{ik}$ is the synaptic weight and $\hat{\mu}_i^{\rm ext}$ and $\hat{\Sigma}_{ij}^{\rm ext}$ are the mean and covariance of an external current, respectively.  
Note that from equation (\ref{eq:sum_cov}), it becomes evident that the synaptic current is correlated even if the presynaptic spike trains are not. 
Next, the first- and second-order moments of the synaptic current are mapped to that of the spiking activity of the post-synaptic neurons. For the LIF neuron model, this mapping can be obtained in closed form through a mathematical technique known as the diffusion approximation~\cite{Feng2006, Lu2010} as
\begin{numcases}{}
\mu_i = \phi_\mu(\bar{\mu}_i,\bar{\sigma}_i),\label{eq:ma_mu}\\
\sigma_i = \phi_\sigma(\bar{\mu}_i,\bar{\sigma}_i),\label{eq:ma_sig}\\
\rho_{ij} = \chi(\bar{\mu}_i,\bar{\sigma}_i)\chi(\bar{\mu}_j,\bar{\sigma}_j)\bar{\rho}_{ij},\label{eq:ma_chi}
\end{numcases}
where the correlation coefficient $\rho_{ij}$ is related to the covariance as $\Sigma_{ij}=\sigma_i\sigma_j\rho_{ij}$. The mapping given by equation (\ref{eq:ma_mu})-(\ref{eq:ma_chi}) is called the moment activation, which is differentiable so that gradient-based learning algorithms can be implemented and the learning framework is known as the moment neural network (MNN)~\cite{Qi2023}.

\subsection{The motion direction detection task}
\subsubsection{Model setup for training}
We trained the MNN model for motion direction detection as detailed in our previous work~\cite{Qi2023}.
The network configuration includes 1054 sensory neurons, 1054 hidden neurons, and 2 readout neurons. 
In the input layer, 527 pairs of intensity and change detectors are evenly distributed across a unit hexagonal grid. 
The sensory neurons, characterized by input response coefficients \(\alpha = \beta = 1\), the temporal angular frequency of \(\omega = 1\), the observation window  \(\Delta t = 1\) and the spatial frequency length \(k = |\mathbf{k}| = 5\pi\), follow our proposed encoding scheme  [equation~(\ref{eq:vo_input_mean})-(\ref{eq:vo_input_cov})] for generating inputs.
The hidden layer comprises a synaptic summation layer, succeeded by a moment batch normalization layer and a moment activation layer. 
After training, the parameters of the moment batch normalization layer were integrated into the synaptic summation layer for optimization.

For effective training of the model parameters \(\bm{\Phi}\), we introduced a loss function based on the readout mean \(\hat{\mu}\) and covariance \(\hat{\Sigma}\),
\begin{equation}
    L(\bm{\Phi}) = \int p(\mathbf{x|\bm{\Phi}}) \arccos\left( \frac{\mathbf{x} \cdot \mathbf{t}}{|\mathbf{x}||\mathbf{t}|}\right) d\mathbf{x} = \mathbb{E}\left[\arccos\left( \frac{\mathbf{x} \cdot \mathbf{t}}{|\mathbf{x}||\mathbf{t}|}\right) \right],
    \label{eq:loss_function}
\end{equation}
where \(\mathbf{x} \sim \mathbb{N}(\hat{\mu}, \hat{\Sigma}|\bm{\Phi})\) represents a Gaussian-distributed readout and \(\mathbf{t}\) is the coordinates of ground truth (\(\cos \theta, \sin \theta\)).
Due to the complexity of this loss function, we approximated it using a finite-sample method. 
Specifically, we estimated \(L(\bm{\Phi})\) as \(\frac{1}{N}\sum_{n=1}^N \arccos (\mathbf{x}^n\mathbf{t}/|\mathbf{x}^n||\mathbf{t}|)\), where each \(\mathbf{x}^n\) is generated using Cholesky decomposition \(\Sigma =  LL^T\), express as \(\mathbf{x}^n = L\mathbf{z}^n + \mathbf{\mu}\) with \(\mathbf{z}^n\) being an uncorrelated unit normal random variable.

The model was trained for 150 epochs using the AdamW optimizer with its default hyperparameters (0.001 learning rate and 0.01 weight decay). 
During each epoch, 10,000 samples were created by randomly selecting contrasts (from 0 to 0.8) and motion directions (from \(-\pi\) to \(\pi\)) to guarantee a comprehensive and varied training dataset.

\subsubsection{Model setup for SNN simulation}
The trained MNN parameters were directly utilized to reconstruct the SNN, as detailed in our previous work~\cite{Qi2023}. For the motion direction detection task, we maintained a consistent contrast level \(c = 0.8\) and selected 50 motion directions uniformly distributed between \(-\pi\) and \(\pi\).
The membrane potentials of hidden neurons were initially set to random values between the resting potential\(V_{\rm res}\) and the threshold  \(V_{\rm th}\). 
To ensure accurate calculation of mean spike rates and trial-to-trial covariance for decoding, we conducted 500 trials for each direction, simulating each for 1256 ms at a time increment of \(\delta t = 0.1\) ms. 
The task inputs were generated via the inhomogeneous Poisson process as described in section \ref{sec:vo_encode}.

We collected responses from the hidden neurons and readouts from the networks at each time step for analysis. 
It is crucial to differentiate these SNN readouts from those obtained from the MNNs. 
The SNN readouts were accumulated over time, and we calculated the discrepancy between the estimated direction represented by these readouts and the actual motion directions.
\subsubsection{Population activity analysis}
We began by tallying the spikes of all sensory and hidden neurons within 1 ms intervals throughout the duration of the simulation. 
This produced a vector representing the population firing rate at each time point. 
The relative deviation of the population spike count was calculated as:
\begin{equation}
D(t) = \frac{R(t) - \langle R \rangle}{\langle R \rangle}, 
\end{equation}
where \(R(t)\) is the population spike count and \(\langle R \rangle\) is its trial-averaged mean rate. 
Using the relative deviation \(D(t)\), we computed the power spectrum density and autocorrelation (as shown in Fig.~\ref{fig:visual_orientation_snn}\textbf{c}), averaging their mean and standard deviation across 50 directions, each estimated with 500 trials.

For further analysis, we used the firing rates of sensory neurons and membrane potentials of hidden neurons recorded from 100 ms to 220 ms in the simulations. 
Data preprocessing involved subtracting the mean value from each trial. 
The Hilbert transform was then applied to calculate the instantaneous phase (\(\psi_i(t)\)) of each neuron. 
The average phase \(\bar \psi(t)\) and the instantaneous Kuramoto order parameter \(r(t)\) determined as follows:
\begin{equation}
r(t) e^{i\bar\psi(t)} = \frac{1}{N}\sum^{N}_{j=1} e^{i \psi_j(t)}. 
\end{equation}
The mean of \(r(t)\) was taken as a measure of synchrony during the simulation.
Since sensory neurons exhibited consistent firing rates in all trials, only the average direction was used for visualization. 
For hidden neurons, the mean and standard deviation of the average phase and the Kuramoto order parameter were first estimated through 500 trials for each direction and then averaged over all 50 directions.
\subsubsection{Information-theoretic analysis}
The mutual information between the motion direction \(\theta\) and the readout \(\mathbf{x}\) is defined in terms of the entropy of all possible responses and conditional responses as
\begin{equation}
    I_{\rm tot}(\mathbf{x}; \theta) = h(\mathbf{x}) - h(\mathbf{x}|\theta).
\end{equation}
The entropy \(h(\mathbf{x})\) and the conditional entropy \(h(\mathbf{x}| \theta)\) are given by
\begin{align}
    h(\mathbf{x}) &= - \int p(\mathbf{x}) \log p(\mathbf{x}) d\mathbf{x},\\
    h(\mathbf{x}|\theta) &= -\sum_i p(\theta_i) \int p(\mathbf{x}|\theta_i) \log p(\mathbf{x}|\theta_i) d\mathbf{x},
\end{align}
where \( p(\mathbf{x}|\theta)\) is assumed to be a Gaussian distribution with readout mean \(\hat{\mathbf{\mu}}\) and covariance \(\hat{\Sigma}\) condition on the stimulus \(\theta\).
The mean \(\hat{\mu}\) covariance \(\hat{\Sigma}\) of the readout \(\mathbf{x}\) within readout time \(\Delta t\) were estimated through accumulative readouts, using data from SNN simulations.  
Assuming a uniform prior and discretizing motion directions into 50 bins (matching the number of directions used for SNN simulations), the distribution of overall neural responses was calculated as follows:
\begin{equation}
    p(\mathbf{x}) = \sum_i p(\theta_i)p(\mathbf{x}|\theta_i).
\end{equation}
We used the information breakdown analysis~\cite{Pola2003, montani2007role} to further dissect the mutual information \(I_{\rm tot}\) into three components, allowing us to assess the contributions of individual neuron responses \(I_{\rm lin}\), signal similarity among neurons \(I_{\rm sigsim}\), and the correlation in neuron activities \(I_{\rm cor}\).
\(I_{\rm lin}\) measures the total amount of information that would be transmitted if all neurons were independent, which is given by:
\begin{equation}
    I_{\rm lin} = \sum_j [h(x_j) - h(x_j | \theta)],
\end{equation}
where \(x_j\) is the possible readout of neuron \(j\).
\(I_{\rm sigsim}\) quantifies the information loss arising from redundancy due to overlap in the tuning curves about response \(x_j\), which is given by:
\begin{equation}
    I_{\rm sigsim} = h(\mathbf{x}_{\rm ind}) - \sum_j h(x_j),
\end{equation}
where the independent population response \(\mathbf{x}_{\rm ind}\) is defined as:
\begin{equation}
    p(\mathbf{x}_{\rm ind} | \theta) = \prod_i p(x_i | \theta).
\end{equation}
The last component \(I_{\rm cor}\) accounts for the rest part of \(I_{\rm tot}\), quantifies the total amount of information due to the correlated activity in the neural responses:
\begin{equation}
    I_{\rm cor} = I_{\rm tot} - I_{\rm lin} - I_{\rm sigsim}
\end{equation} 
Since there is no analytical expression for the entropy of unconditional neural response \(p(x)\), we estimated this information through Monte Carlo sampling.
The mean \(\hat{\mathbf{\mu}}\) and covariance \(\hat{\Sigma}\) were computed based on accumulative readouts for all directions in a readout time window \(\Delta t\) (Fig~\ref{fig:visual_orientation_snn}\textbf{e}). 
Subsequently, we sampled 10000 data points based on these \(\hat{\mu}\) and \(\hat{\Sigma}\) at time \(t\) to empirically estimate the information as a function of the readout time window \(\Delta t\).
The theoretic bound of \(I_{\rm tot}\) is the entropy of direction, where \(H(\theta) = \sum_i p(\theta_i)\log \theta_i = \log 50\) nats and we defined the time when the readout information converges as the first time when \(|I_{\rm tot} - H(\theta)| < 0.1\) nats.

\subsection{The fine-grain classification task}
\subsubsection{Model setup for training}
For our fine-grain classification task, we utilized the Caltech-UCSD Birds-200-2011 dataset \cite{WahCUB_200_2011}, comprising 11,788 images of 200 bird species. 
Adhering to the recommended dataset split, we divided it into a training set of 5,994 images and a test set of 5,794 images.

Our image preprocessing was in agreement with previous work~\cite{Wang2020}. 
We resized the images to have a shorter side of 448 pixels, then center-cropped them to \(448 \times 448\) resolution. 
To augment the training data, the images were randomly flipped horizontally. Normalization was applied using z-score normalization, with RGB channel means (0.485, 0.456, 0.406) and standard deviations (0.229, 0.224, 0.225) derived from the ImageNet dataset~\cite{simonyan2014very}.

Feature maps were extracted using a pretrained VGG16 model~\cite{simonyan2014very}  from PyTorch, with its parameters fixed. The resulting feature maps, of dimensionality \(512 \times 28 \times 28\), were flattened and processed to calculate channel-wise mean and covariance [equation~(\ref{eq:theory_define_mean})-(\ref{eq:theory_define_cov})], serving as inputs for our MNN model.

To evaluate our theory, we modified the inputs in various ways. 
The \textit{With corr} model received both mean and covariance as inputs, while the \textit{Without corr} model had its covariance off-diagonal entries zeroed, erasing correlation information. 
The \textit{ANN} was trained using only the mean. Each model, including the MNNs and ANN, consisted of an input layer (512 neurons), two hidden layers (1024 neurons each), and a readout layer (200 neurons). 
They were trained 5 times, each time for 150 epochs, using standard cross-entropy loss and AdamW optimizer with the default setting.

\subsubsection{Model setup for SNN simulation}
We reconstructed the SNNs using the parameters of the trained MNNs under both \textit{With corr} and \textit{Without corr} conditions. 
To evaluate the performance of the SNNs on the fine-grain classification task, we simulated the network for each image in the test set, each for 100 trials, with each simulation running for 150 ms at a time increment of \(\delta t = 0.1\) ms. 
The task inputs were generated by sampling from a Gaussian distribution, using the mean and covariance of the respective input conditions. 

During these simulations, we accumulated the readouts and determined the prediction of the model by identifying the dimension index with the highest cumulative readouts. 
The accuracy of the prediction was determined by taking the average of the proportion of correct predictions made for all the samples in the test set in all the trials.

\section*{Data Availability}
The Caltech-UCSD Birds-200-2011 dataset and the pre-trained VGG16 model are publicly available from \href{https://www.vision.caltech.edu/datasets/cub_200_2011/}{https://www.vision.caltech.edu/datasets/cub\_200\_2011/} and \\ \href{https://pytorch.org/vision/stable/models/vgg.html}{https://pytorch.org/vision/stable/models/vgg.html} respectively.

\section*{Code Availability}
The code used in this article is available at \href{https://github.com/BrainsoupFactory/moment-neural-network}{https://github.com/BrainsoupFactory/moment-neural-network}.

\section*{Acknowledgments}
This work is jointly supported by Shanghai Municipal Science and Technology Major Project (No.2018SHZDZX01), the National Natural Science Foundation of China (No. 62072111), ZJ Lab and Shanghai Center for Brain Science and Brain-Inspired Technology.

\section*{Author Contributions}
Conceptualization: Z.Z., Y.Q. and J.F.; 
Methodology: Z.Z., Y.Q., W.L., and J.F.;
Investigation: Z.Z. and Y.Q.;
Software: Z.Z. and Y.Q.;
Visualization: Z.Z. and Y.Q.;
Writing — original draft: Z.Z. and Y.Q.;
Writing — review \& editing: Z.Z., Y.Q. , W.L., and J.F.;
Supervision: J.F.

\section*{Competing interests}
The authors declare no competing interests.

\section*{Materials \& Correspondence}
Correspondence and requests for materials should be addressed to J.F.
\newpage
\bibliography{reference}
\bibliographystyle{naturemag}
\setcounter{figure}{0}
\renewcommand{\figurename}{Extended Data Fig.}
\begin{figure}
    \includegraphics[width=\linewidth]{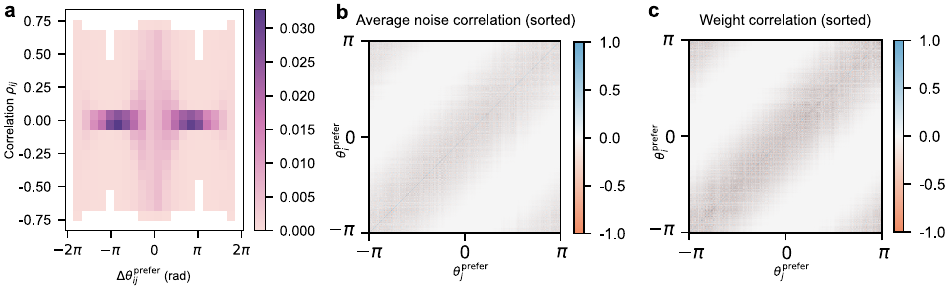}
    \caption{\textbf{Analysis of noise correlation among the hidden neurons under the high contrast condition.}
    \textbf{a.} Distribution of noise correlation \(\rho_{ij}\) for pairs of hidden neurons (\(i\neq j\)) plotted against the difference in their preferred directions (\(\Delta\theta_{ij}^{\rm prefer} = \theta_i^{\rm prefer} - \theta_j^{\rm prefer}\)). The color bar represents the joint probability of \(\Delta\theta_{ij}^{\rm prefer}\) and \(\rho_{ij}\).
    \textbf{b.} Average noise correlation among hidden neurons as motion directions vary. 
    \textbf{c.} Correlation patterns in the trained connection strengths \(W_{\rm in}\) from all sensory neurons to individual hidden neurons.
    In both \textbf{b} and \textbf{c}, hidden neuron indices are sorted based on their preferred directions, ranging from \(-\pi\) to \(\pi\), and color bars indicate the strength of correlation coefficients. 
    All results are obtained at the high contrast level where \(c = 0.8\).
    \label{fig_s1: correlation_vs_preferred_direction}}
\end{figure}

\end{document}